\documentclass[runningheads]{llncs}
\usepackage{graphicx}
\usepackage[colorlinks]{hyperref}
\usepackage{verbatim}
\usepackage{multirow}

\begin{document}
\title{DAT: Data Architecture Modeling Tool for Data-Driven Applications}

\author{Moamin Abughazala\inst{1} \orcidID{0000-0003-4946-6269} \and
Henry Muccini\inst{1} \orcidID{0000-0001-6365-6515} \and Mohammad Sharaf\inst{2} }
\authorrunning{Abughazala}

\institute{University of L'Aquila, L'Aquila, Italy\\
\email{moamin.abughazala@graduate.univaq.it}\\ \email{henry.muccini@univaq.it}
\and
An Najah N. University, Nablus, Palestine\\  \email{sharaf@najah.edu}
}

\maketitle              

\begin{abstract}
Data is the key to success for any Data-Driven Organization, and managing it is considered the most challenging task. Data Architecture (DA) focuses on describing, collecting, storing, processing, and analyzing the data to meet business needs.
In this tool demo paper, we present the DAT, a model-driven engineering tool enabling data architects, data engineers, and other stakeholders to describe how data flows through the system and provides a blueprint for managing data that saves time and effort dedicated to Data Architectures for IoT applications.
We evaluated this work by modeling five case studies, receiving expressiveness and ease of use feedback from two companies, more than six researchers, and eighteen undergraduate students from the software architecture course.

\end{abstract}
\section{Introduction}

The International Data Corporation (IDC) \cite{idc} expects that by 2025 there will be more than 175 zettabytes of valuable data for a compounded annual growth rate of 61\%. Ninety zettabytes of data will be from IoT devices, and 30\% of the data generated will be consumed in real-time. 
A {\em data architecture} is an integrated set of specification artifacts used to define data requirements, guide integration, control data assets, and align data investments with business strategy. It also includes an integrated collection of master blueprints at different levels of abstraction \cite{10.5555/3165209}.

This tool demo paper presents the {\em Data Architecture Modeling Tool (DAT)}, an architecture modeling tool for the model-driven engineering of data architecture for data-driven applications.  

DAT (Data Architecture Modeling Tool) is a data architecture modeling tool for IoT applications that shows how data flows through the system and provides a blueprint for it. It allows the stakeholders to describe two levels of data architecture: high-level Architecture (HLA) and Low-Level Architecture (LLA). It focuses on representing the data from source to destination and shows formats, processing types, storage, analysis types, and how to consume it.

The rest of this tool demo paper is organized as follows. The methodology is presented in Section 2. The application of DAT to a real case study is described in Section 3. The DAT evaluation is presented in Section 4. Related work is discussed in Section 5, while conclusions are drawn in Section 6. 
\section{Background}
\label{sec:Background}

The main focus of this paper is to describe the data architecture of IoT applications through the {\em Data Modeling Language (DAML)}. Section \ref{42010} shows ISO/IEC/IEEE 42010:2011 standard. CAPS Framework in Section \ref{caps}. Section \ref{DATTool} shows DAML and reports on the technologies used to implement the DAT. 

\subsection{IEEE/ISO/IEC 42010 Architecture Description}
\label{42010}
\begin{center}
 \begin{figure*}[ht!]
	\centering
	{
	    	\includegraphics[width=0.9\textwidth]{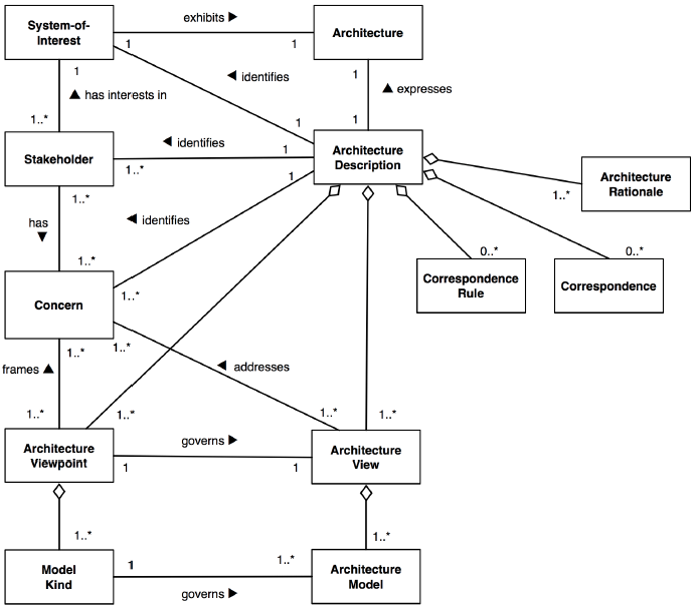}
	}
	\caption{Content model of an architecture description (ISO/IEC/IEEE 42010)}
	\label{fig:IEEE_ISO_42100}
\end{figure*}
\end{center}

Our work is built on the conceptual foundations of the ISO/IEC/IEEE 42010:2011, \textit{Systems and software engineering --- Architecture description}~\cite{42010} standard, to investigate the essential elements of data architecture description for IoT applications. The standard handles architecture description (AD), the practices of recording software, system, and enterprise architectures so that architectures can be understood, documented, analyzed, and realized. Architecture descriptions can take many forms, from informal to carefully specified models. \newline
The content model for an architecture description is illustrated in Figure~\ref{fig:IEEE_ISO_42100}. The {\em Architecture viewpoint} is a fundamental building block representing common ways of expressing recurring architectural concerns reusable across projects and organizations. It encapsulates {\em model kinds} framing particular {\em concerns} for a specific audience of system {\em stakeholders}. The concerns determine what the model kinds must be able to express: e.g., security, reliability, cost, etc. A model determines the notations, conventions, methods, and techniques. Viewpoints, defining the contents of each architecture {\em view}, are built up from one or more model kinds and {\em correspondence rules}, linking them together to maintain consistency.

\subsection{The CAPS Modeling Framework}
\label{caps}
CAPS \cite{muccini2017caps} is an environment where Situational Aware Cyber-Physical Systems (SiA-CPS) can be described through software, hardware, and physical space models. The CAPS found three main architectural viewpoints of extreme importance when describing a
SiA-CPS: the software architecture structural and behavioral view (SAML), the hardware view (HWML), and the physical space view
(SPML).

This environment is composed of the CAPS modeling framework\footnote{CAPS: http://caps.disim.univaq.it/} and the CAPS code generation framework \cite{sharaf2017architecture} \cite{sharaf2018arduino} that aim to support the architecture description, reasoning, design decision process, and evaluation of the CAPS architecture in terms of data traffic load, battery level and energy consumption of its nodes.

\subsection{The Important of Data Architecture}\label{DATTool}
Data architecture is important because it helps organizations manage and use their data effectively. Some specific reasons why data architecture is important to include: 
\begin{enumerate}
\item  Data quality: it helps to collect, store, and use data consistently and accurately. This is important for maintaining data integrity and reliability and avoiding errors or inconsistencies impacting business operations.

\item Data security: it helps to protect data from unauthorized access or modification and ensure that it is used compliantly and ethically. This is particularly important in industries with strict regulations, such as healthcare or finance.

\item Organizational efficiency: it helps organizations better understand and manage their data, increasing efficiency and productivity. By defining the structures, policies, and standards that govern data within an organization, data architecture can help streamline processes and improve decision-making.

\item Business intelligence and analytics: it is essential for organizations to collect, store, and analyze large amounts of data. This can support better decision-making, improve customer relationships, and drive business growth.

\item Scalability and flexibility: A well-designed data architecture can support the growth and evolution of an organization. It allows organizations to easily add new data sources, incorporate new technologies, and adapt to changing business needs.
\end{enumerate}

\subsection{The DAT Tool}\label{DATTool}
The DAT modeling framework \footnote{DAT Tool Source Code can be found at \url{https://github.com/moamina/DAT}} \footnote{DAT Tool video demo: \url{https://youtu.be/Du0VDg1CLlQ}} gives data architects the ability to define a {\em data view} for data-driven IoT Applications through the DAML modeling language \cite{10092710}. 
\begin{center} 
   \begin{figure*}[h!]
	\centering
	\caption{The Data View of CAPS}
	{
	    	\includegraphics[width=1\textwidth]{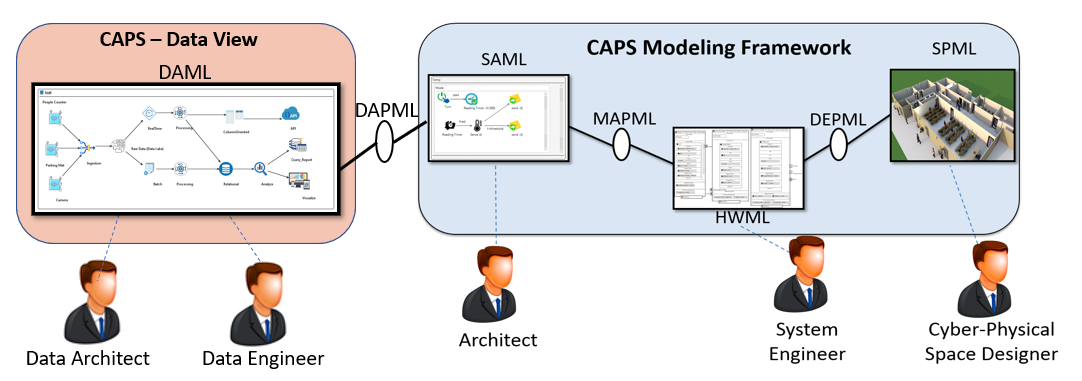}
	}
	\label{fig:dataView_CAPS}
    \end{figure*}
\end{center}
\subsubsection{Technologies.}
Our tool is based on MDE. For that, we use Eclipse Modeling Framework (EMF) \cite{emf}  for building tools and applications based on the structure data model, which consists of three main parts, EMF Core, includes a meta-model for describing the models. EMF Editors contains generic reusable classes for building editors for EMF models. Eclipse Epsilon \cite{epsilon} is a Java-based scripting language for model-based software engineering tasks (e.g., model-to-model transformation and model validation) which strongly support EMF and works with UML, XML, Simulink, etc. To create graphical editors and views for the EMF models, we used Eugenia \cite{eugenia}. It is a tool to create a graphical model editor by generating the .gmfgraph, .gmftool, and .gmfmap models that the GMF editor from a single annotated Ecore meta-model needs. 

DAT implements the DAML meta-model to be considered a fourth view (data view) for the CAPS, as shown in Figure \ref{fig:dataView_CAPS}. How the DAT supports the modeling of data views and its application to actual use cases will be presented in Section 3.
\subsubsection{Methodology.} DAT is built based on a meta-model containing a data architecture as a top root meta-class. Any \textbf{data architecture} of IoT can contain a set of \textbf{DataNodes} (components) and \textbf{connections}. A Component is considered a computational unit with an internal state and a known interface \cite{component}. Data nodes can interact by passing data through \textbf{data ports}. A component's internal state is denoted by the current behavior of data representation and its values. Data representation includes \textit{data formats, storage technologies, location, and processing type}. Every Node Behavior has a set of behavioral elements denoted by actions and events that depict the data flow within the component. This element can be executed when a previous action in the behavioral data flow has been achieved or triggered by an event like \textbf{ReceiveData}. Other main actions are \textbf{Generation}, \textbf{Ingestion}, \textbf{Process},  \textbf{Store}, \textbf{Analyze}, and \textbf{Consume}. An \textbf{event} is triggered in response to the external stimulus of the component. To show the data flow and connection between the events and actions, we use \textbf{links}.

\subsubsection{Steps To Use.}
The architect needs to follow the following steps to use the tool to model any case:
\begin{enumerate}
\item Download the source code from The GitHub \footnote{DAT Tool Source Code can be found at \url{https://github.com/moamina/DAT}} and follow the steps in the tool demo video to lunch the tool \footnote{DAT Tool video demo: \url{https://youtu.be/Du0VDg1CLlQ}}.

\item Define which level of abstraction you need (High-Level or Low-Level). You could use a single Data Node at the High-Level, whereas you need to define the structure and behavior at the low level. 

\item Define the main data nodes and the connection between them to determine the order of each node, such as the data source is the first data node. Ingestion comes the second one, and the connection shows the data flow from the source to the ingestion data node.

\item At the internal behavior, you could use data elements (low-level elements), such as data formats (JSON, XML, Video, ...), and sub-operation, such as (classification, data reduction, cleaning, validation, filter, classification, ...).
\end{enumerate}
\section{Real Use Cases}
This section introduces the existing data architecture description used by three companies contributing to the DAT tool. We have chosen three cases from five cases to present our tool in the tool paper.

\subsection{Operational Data Warehouse}
The data warehouse (DW) depends on data from different sources within the operational company system. These data sources can be data from RDS (MySQL Relational-DB), documents-based data in DynamoDB, and real-time data streams. The DW has data batching mechanisms that perform (complete data Extract, Transform, and Load (ETL)) processes on these data sources to load into the final DW tables and data models for reporting purposes. The ETL process is built through data batches using large-scale data processing and a file system (e.g., AWS S3). Batches run in an hourly-based fashion. The final DW data model is saved in Column-oriented format using AWS S3. 

Data from RDS will wait a specific time to be extracted, transformed, and saved in the column-oriented format on file system technology (AWS S3).
For the data that comes from DynamoDB streams or real-time data streams, financial details will be added to part of this data for reporting purposes. Then this data is sent to the ingestion stage, extracted, transformed into a column-oriented format, and stored on file system technology (AWS S3) that will be consumed later by the batches to be processed and saved in the final tables. 

The ETL batches will check for the new files on the staging tables; whenever a new file is found, the batch will extract the data, transform it, and save the related final tables on the final DW. Once the data is ready in the final tables, it will be ready to be queried for reporting and data export purposes. 

Data-consuming micro-services use a query engine (Presto) to query the data in the DW. DW consumers could be reports, dashboards, or others.
Report generator micro-services provide all the reports and dashboards with data. The warehouse exporter is responsible for creating data CSV exported files based on specific data templates and sending them to external customer endpoints such as(SFTP, FTP, S3, and emails).
Reporting management, the purpose of this service is to manage and maintain the end user's custom configurations and settings of their
preference in the dashboards and reports layouts. 
Tagging management, this service is built for a specific custom report (Operational Flash Dash) that gives the end user ability to decide on and design his report hierarchy and data drill down from the manager position perspective.

\subsection{Hydre}
\begin{center} 
   \begin{figure*}[!t]
	\centering
	\makebox[\textwidth]
	{
	    	\includegraphics[width=1\textwidth]{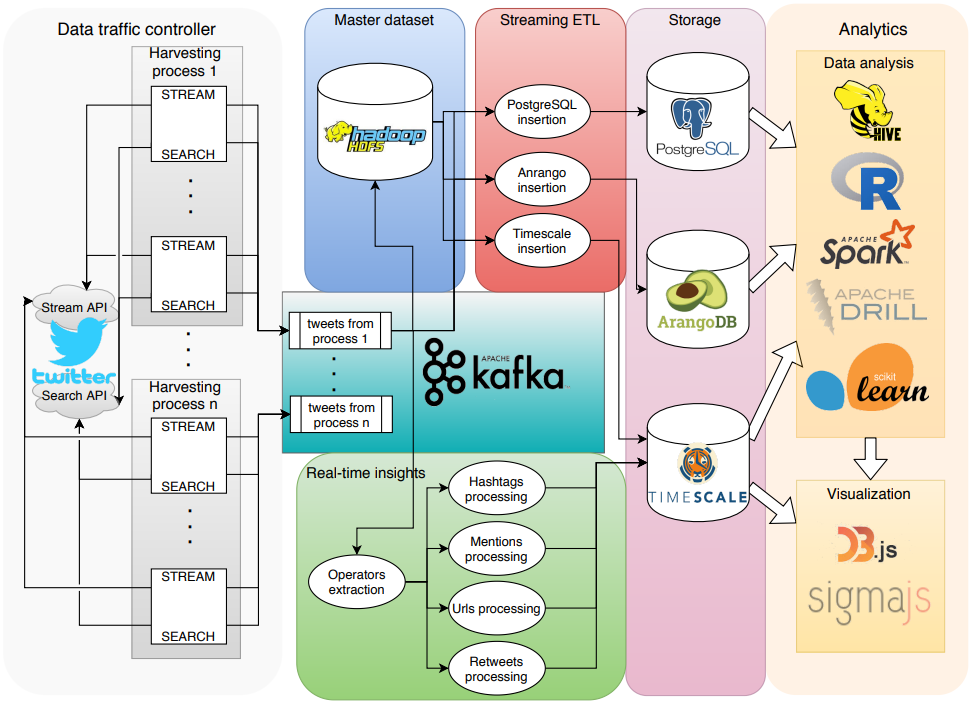}
	}
	\caption{The Hydre architecture}
	\label{fig:Hydre_realCase}
    \end{figure*}
\end{center}

This case ~\ref{fig:Hydre_realCase} from Lambda+ paper's author \cite{gillet2021lambda+}, is from a (Cocktail) research project which aims to study the discourses in two domains in health and food, as well as to identify weak signals in real-time using social network data. The data come from Twitter, compute real-time insights and store data for exploratory analysis.

The case contains other components. The master dataset is implemented with file system technology (Hadoop HDFS). Raw data (tweets) are stored as lines of files, and data re-processing can be done by reading  and sending each line as is in another Kafka topic. The streaming ETL uses Kafka consumers to insert data in the micro-batch. The streaming ETL applies transformations and then stores tweets in the storage component. That includes relational, graph, and time series DBMSs. These databases are used for exploratory analyzes, mainly performed with Jupyter notebooks. Alongside, the real-time insights component extracts and aggregates several information about the harvesting, such as popular hashtags or users, using Kafka Streams. It stores the results in the time series database. Although this insertion is a side-effect of the stream processing, it is an idempotent action because the count of the elements will always yield the same result with an effective once guarantee. This result is stored for each element, replacing the old value if it already exists.

\subsection{Errors Data Pipeline}
This case shows the data pipeline for data errors from different printers. The error data text files come from other printers in JSON format. The data represent the error that happened in different printers, and the data could be the version of the printer, location, ink type, software version, time, etc. the data will be sent to AWS S3 and saved in the same format. The customer could see the raw error data using a query engine. The data will be processed and transferred into helpful information after a specific time, then converted to parquet format (Column-oriented). After that, the data will be transformed to CSV format and then to relation database format to be ready for query from the customer.
\section{Evaluation}
The DAT cases were evaluated through interviews with seven industry professionals from two companies of different domains and different maturity levels and one external researcher; table \ref{tab:EvalTbl} shows more about the roles of the evaluators. The evaluation section will be structured in terms of agreements and suggestions for improvement. 
\label{sec:evaluation}
\begin{table}[h!]
	\centering
	\caption{Outline of use cases and roles of the evaluators}
	\label{tab:EvalTbl}
	\begin{tabular}{|p{2.5cm}|p{6cm}|p{4cm}|} 
	\hline
	\textbf{Company} & \textbf{Use cases} & \textbf{Experts Roles} \\
	\hline
	\multirow{4}{*}{\textbf{Company A}} & \multirow{2}{*}{Operational Data Warehouse} & Big Data Team Lead \\
	\cline{3-3}
	& & Big Data Architect\\
	\cline{2-3}
	& \multirow{2}{*}{Analytical Data Warehouse} & Big Data Engineer\\
	\cline{3-3}
	&  & Big Data Architect\\
	\hline
	
	\multirow{3}{*}{\textbf{Company B}} & \multirow{3}{*}{Data Pipeline} & Big Data Team Lead \\
	\cline{3-3}
	& & Big Data Architect\\
	\cline{3-3}
	&  & Big Data Engineer\\
	\hline
	
	\textbf{Researchers} & Data Architectures(Lambda, Kappa) & Researcher \\
	\hline
	
	\textbf{Locally} & NdR Data Architecture & Students \\
	\hline
	\end{tabular}
\end{table}

\subsection{Errors Data Pipeline}
\begin{center} 
   \begin{figure*}[!h]
	\centering
	\makebox[\textwidth]
	{
	    	\includegraphics[width=1\textwidth]{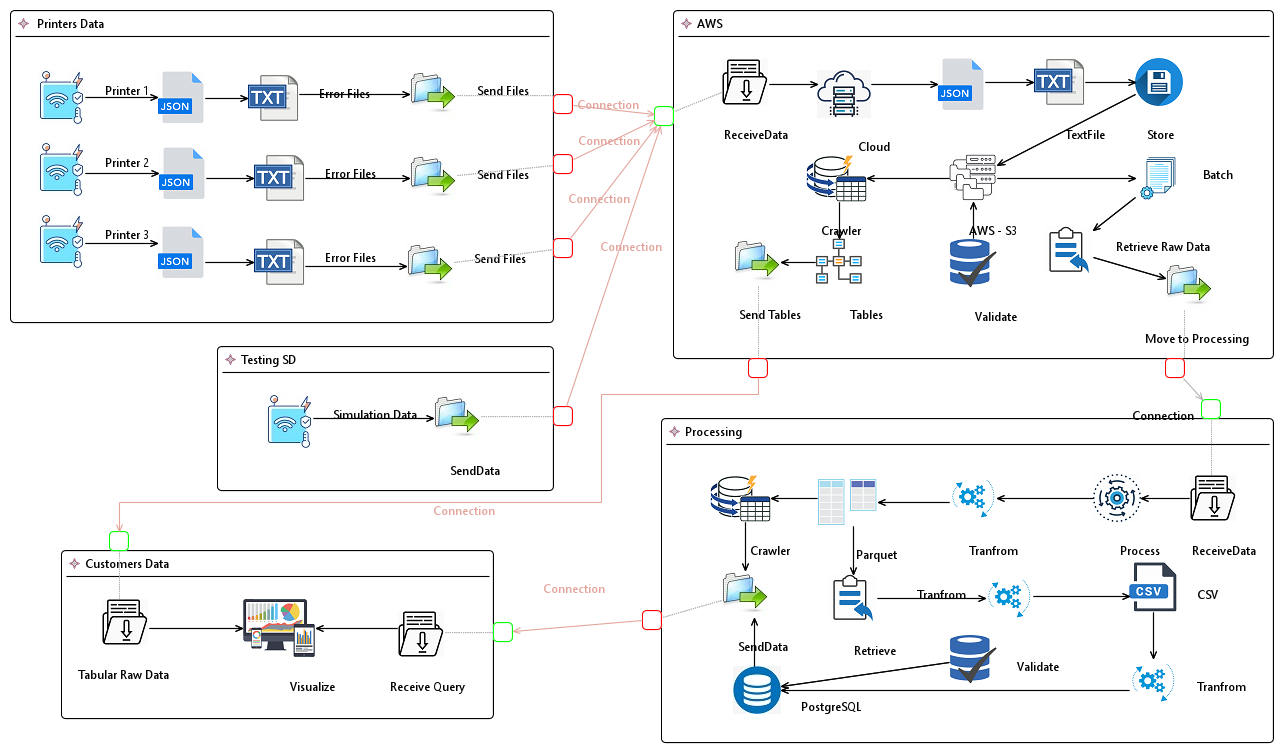}
	}
	\caption{Errors Data Pipeline}
	\label{fig:hp_err}
    \end{figure*}
\end{center}

Figure \ref{fig:hp_err} shows the Errors Data Pipeline using DAT. The first author presented the models and collected the practitioners' feedback. 

\textbf{Agreements:} The model described the data flow from generation to destination. The model was easy to understand for new data engineers. The tool has the flexibility to change and add new nodes.

\textbf{suggestion:} It is good to include the data quality metrics that could apply to the data at each stage. 

\subsection{Hydre}
\begin{center} 
   \begin{figure*}[!h]
	\centering
	\makebox[\textwidth]
	{
	    	\includegraphics[width=1\textwidth]{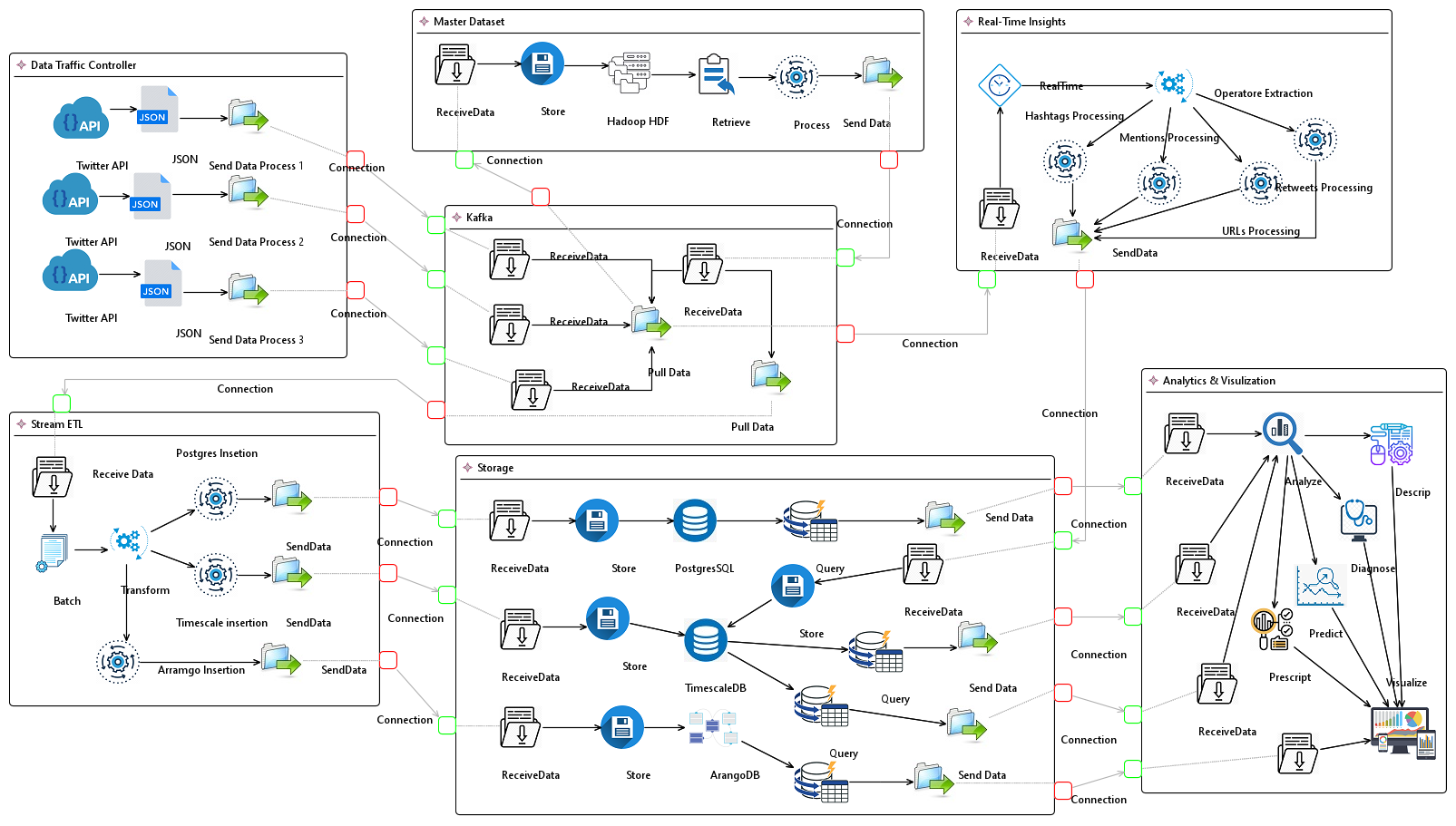}
	}
	\caption{Hydre (Lambda+ Example)}
	\label{fig:Hydre}
    \end{figure*}
\end{center}

Figure \ref{fig:Hydre} shows the Hydre model using DAT. The first author presented the models and collected feedback from Lambda+'s author. 

\textbf{Agreements:} The model represents very well the Hydre case. The indications of when data are stored on disk are helpful, especially when working on a big data architecture with people who don't know each technology's details. The real-time and batch elements are useful as well.

\textbf{suggestion:} the first suggestion is similar to the first case related to data interaction patterns. The second suggestion was a starting point for us to provide two levels of architecture, High-level architecture (HLA) and Low-Level architecture (LLA).
\subsection{Operational Data Warehouse}
\begin{center} 
   \begin{figure*}[!ht]
	\centering
	{
	    	\includegraphics[width=1\textwidth]{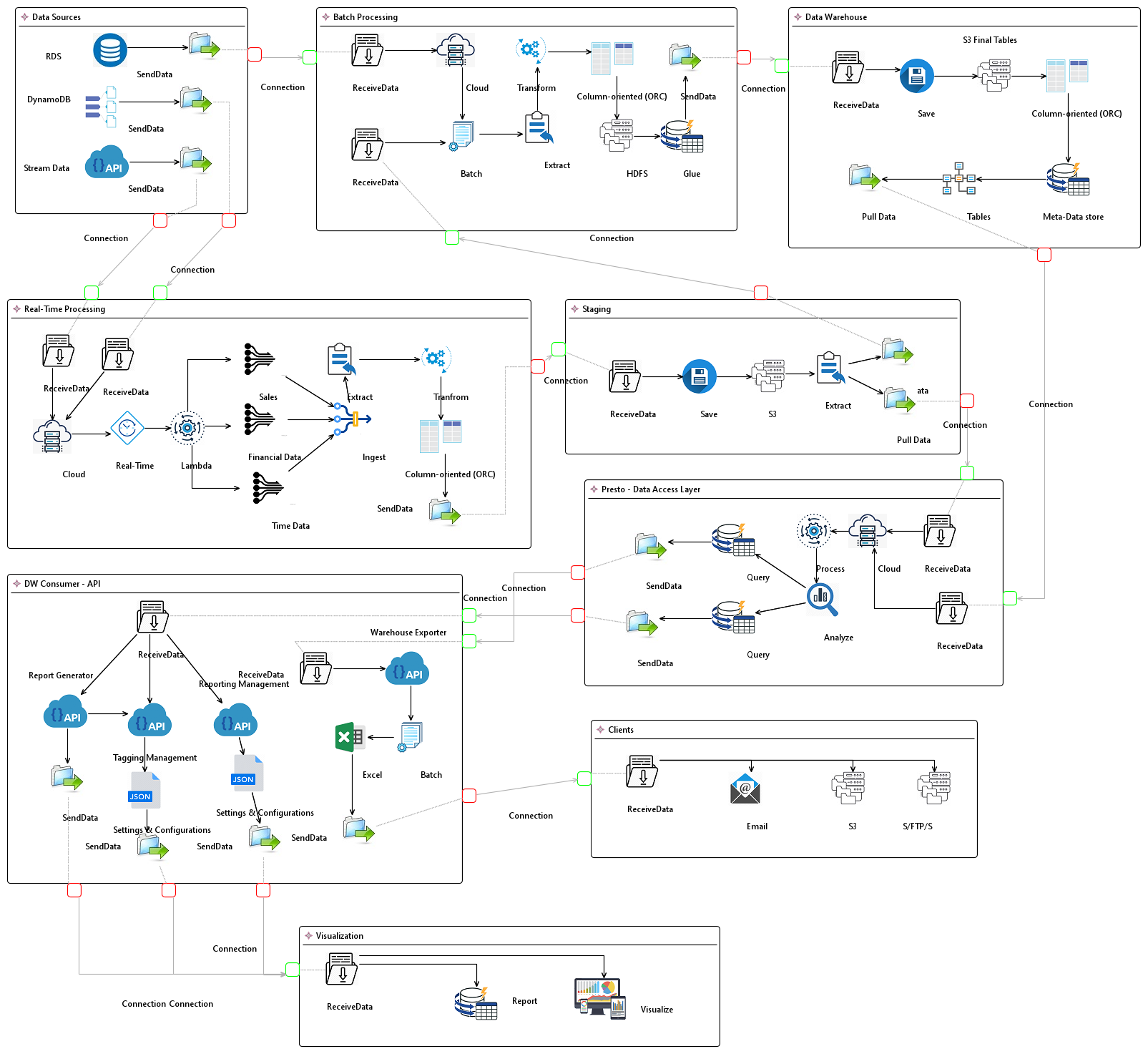}
	}
	\caption{Operational Data Warehouse}
	\label{fig:odw}
    \end{figure*}
\end{center}

Figure \ref{fig:odw} shows the ODW model using DAT. The first author presented the models and collected the practitioners' feedback. 

\textbf{Agreements:} The model was able to describe the details of the case and was easy to share and understand by different teams in other parts of the world. The model was a good communication language between team members, which means easy to avoid misinterpretations.

\textbf{suggestion:} In the current version of the DAT, the only way to show how different components interact with each other is to send/receive data. The suggestion was to include all data interaction patterns (request/response, publish/subscribe, pull/push, and others).


\section{Related Work}

This section reviews relevant studies that are related to 
exploiting the most related research to data-driven IoT. Raj and Bosch \cite{raj2020modelling} proposed a conceptual model for a data pipeline, which contains two main components (nodes and connectors); the node represents the main abstract data node, and the connection represents the way to carry and transmit the data between nodes. The DAT provided two levels of architecture, HLA (High-Level Architecture) and LLA (Low-Level Architecture), which is a more details architecture that gives the ability to model the behavior of each node by describing sub-action, data formats, location, processing type, etc. 
Borelli  \cite{borelli2020architectural}
proposed a classification for main software components and their relationships to model a software architecture for particular IoT applications. These components represent the abstract components. DAT can describe all of the mentioned components and their behavior too. 
Erraissi \cite{erraissi2018data} \cite{erraissi2019big} proposed a meta-model for  data sources, ingestion layers, and Big Data visualization layer. DAT can describe the data in each layer (source or generation, Ingestion, Processing, storing, Analyzing, and Consuming). 
Nesi \cite{nesi2018auditing} provided a solution based on a set of instruments to collect the data in real-time, store, and audit data flow for IoT smart City architecture. DAT is an architecture-driven tool to show how data flow from the source to the final destination at an abstract concept level; it is not a technologies-based tool.

\section{Conclusion and Future Work}

This tool demo paper has presented the DAT, an architecture description, and the associated modeling platform for the model-driven engineering of Data Architecture for IoT. It is implemented on top of f the Eclipse Modeling Framework. It can allow the stack-holders to describe two levels of data architectures, high-level Architecture (HLA) and Low-Level Architecture (LLA).

This is an initial starting point for our future work plan, which can be extended to include finishing the current running evaluations with other companies and trying to model different big data patterns and architectures. Second, try to integrate the DAT with other existing technologies and tools.


\section{Acknowledgment}
The authors would like to thank Prof. Giovanni Stilo, Prof. Annabelle Gillet (Lambda+), Prof. Karthik Vaidhyanathan, Mostafa Shaer, Itay, and Roi from HP Team, Mustafa Tamim and Anas Eid from Harri Team, Mudassir Malik, Apurvanand Sahay, and Arsene Indamutsa as a researcher for their contributions in the evaluation.\newline

\bibliographystyle{splncs04}  
\bibliography{refs}
\end{document}